\documentclass[preprint,12pt]{elsarticle}
\usepackage{amssymb}
\usepackage{graphicx}

\journal{Physics Letters B}

\begin{document}

\title{
%Investigating $\alpha$-particle clustering  
 Probing clustering in excited alpha-conjugate nuclei}

\author[ipno]{B.~Borderie}
\author[ipno,ifin]{Ad. R. Raduta}
\author[ipno]{G.~Ademard}
\author[ipno]{M.~F.~Rivet\fnref{fn1}}
\author[INFNSezioneCatania]{E.~De~Filippo}
\author[INFNSezioneCatania,FisicaCatania,Bologna]{E. Geraci}
\author[ipno,lpc]{N. Le Neindre}
\author[lns]{R.~Alba}
\author[lns]{F. Amorini}
\author[INFNSezioneCatania]{G.~Cardella}
\author[saha]{M.~Chatterjee}
\author[lyon]{D.~Guinet}
\author[lyon]{P.~Lautesse}
\author[INFNSezioneCatania,csf]{E. La Guidara}
\author[lns,kore]{G.~Lanzalone}
\author[INFNSezioneCatania]{G. Lanzano\fnref{fn1}}
\author[lns,Naples]{I. Lombardo}
\author[lpc]{O.~Lopez}
\author[lns]{C.~Maiolino}
\author[INFNSezioneCatania]{A.~Pagano}
\author[INFNSezioneCatania]{M.~Papa}
\author[INFNSezioneCatania]{S.~Pirrone}
\author[INFNSezioneCatania,FisicaCatania]{G. Politi}
\author[lns,FisicaCatania]{F. Porto}
\author[lns,FisicaCatania]{F. Rizzo}
\author[lns,FisicaCatania]{P.~Russotto}
\author[ganil]{J.P.~Wieleczko}
\fntext[fn1]{deceased}

\address[ipno]{Institut de Physique Nucl\'eaire, CNRS/IN2P3,
Univ. Paris-Sud, Universit\'e Paris-Saclay, Orsay, France}
\address[ifin]{National Institute for Physics and Nuclear Engineering,
Bucharest-Magurele, Romania}
\address[INFNSezioneCatania]{INFN, Sezione di Catania, Italy}
\address[FisicaCatania]{Dipartimento di Fisica e Astronomia,
Universit\`a di Catania, Italy}
\address[Bologna]{INFN, Sezione di Bologna and Dipartimento di Fisica,
Universit\`a di Bologna, Italy}
\address[lpc]{LPC, CNRS/IN2P3, Ensicaen, Universit\'{e} de Caen, 
Caen, France}
\address[lns]{INFN, Laboratori Nazionali del Sud, Catania, Italy}
\address[saha]{Saha Institute of Nuclear Physics,
Kolkata, India}
\address[lyon]{Institut de Physique Nucl\'eaire, CNRS/IN2P3,
Univ. Claude Bernard Lyon 1, Universit\'e de Lyon, Villeurbanne, France}
\address[csf]{CSFNSM, Catania, Italy}
\address[kore]{Facolt\'a di Ingegneria e Architettura, Universit\`a Kore, Enna, Italy}
\address[Naples]{Dipartimento di scienze Fisiche,
Universit\`a Federico II and INFN, Sezione di Napoli, Italy}
\address[ganil]{GANIL, (DSM-CEA/CNRS/IN2P3), Caen, France}

\begin{abstract}
The fragmentation of quasi-projectiles from the nuclear reaction
$^{40}$Ca+$^{12}$C at 25 MeV per nucleon bombarding energy was used to produce
$\alpha$-emission sources. From a careful selection of these sources
provided by a complete detection and
from comparisons with models of sequential and simultaneous decays,
evidence in favor of $\alpha$-particle clustering from excited
$^{16}O$, $^{20}Ne$ and $^{24}Mg$ is reported.

\end{abstract}


%\today

\maketitle

%%%%%%%%%%%%%%%%%%%%% Clustering in general
Clustering is a generic phenomenon which can appear in homogeneous matter when density
decreases; the formation of galaxies as well as the disintegration
of hot dilute heavy nuclei into lighter nuclei are extreme examples
occurring in nature.
As far as nuclear physics is concerned, the nucleus viewed as a collection of
$\alpha$-particles was discussed very early~\cite{gamow_1929} and in the last forty years both
theoretical and experimental efforts were devoted to the study of clustering phenomena in
nuclei~\cite{cluster1,cluster2,cluster3}.
It was recently shown clear deviations from statistical models in
the decay pattern of excited $^{24}Mg$ nuclei: measured emission channels
involving multiple $\alpha$-particles are 20 to 40\% more probable
than expected~\cite{Bai13,Mor141,Mor142}.
It is also known that low density
nuclear matter is predicted to be unstable against cluster formation, mainly
$\alpha$-particles~\cite{ropke_nm_prl1998,beyer_plb2000}, and that excited states of
alpha-conjugate nuclei like $^{12}C$ and
$^{16}O$ are well described assuming a weakly interacting gas of
almost free $\alpha$-particles, which can be qualified as an $\alpha$-particle
condensate state~\cite{tohsaki_prl2001,funaki_prl2008}.
Very recently the formation of $\alpha$-particle clustering from
excited expanding alpha-conjugate
nuclei was revealed in two different constrained self-consistent mean
field calculations~\cite{girod_prl2013,ebran_2014}.

%%%%%%%%%% clustering from the exp. perspective

The aim of the present Letter is to search for experimental
evidence of $\alpha$-particle clustering from excited
and consequently expanding
alpha-conjugate nuclei.
The chosen experimental strategy was to use
the reaction $^{40}$Ca+$^{12}$C at an incident energy (25 MeV per
nucleon) high enough to possibly produce some hot expanding reaction
products, associated with a
high granularity, high solid angle particle array (to precisely reconstruct
directions of velocity vectors). Then, by selecting the appropriate
reaction mechanism and specific events, 
the required information was inferred.

%%%%%%%% our data %%%%%%%%%%%%%%%%%%%%%%%%%%%%%%%%%%%%%%%%%%%%%ù

The experiment was performed at INFN,
Laboratori Nazionali del Sud in Catania, Italy. 
The beam, impinging on a thin carbon target (320 $\mu$g/cm$^2$),
was delivered by the Superconducting Cyclotron and the
charged reaction products were detected by the CHIMERA 4$\pi$
multi-detector~\cite{chimera}. The beam intensity was kept around
$10^7$ ions/s to avoid pile-up events. Events were registered when the
silicon detectors of at least three telescopes were fired.
CHIMERA consists of 1192 telescopes made of $\Delta E$ silicon detectors
200-300 $\mu$m thick (depending on polar angle) and CsI(Tl) stopping
detectors. They are mounted on 35
rings covering 94\% of the solid angle, with polar angle ranging from
1$^{\circ}$ to 176$^{\circ}$. 
The solid angle corresponding to each module  
varies between 0.13 msr
at forward angles and 35.4 msr at the most backward angles. 
Among the most interesting characteristics of CHIMERA are 
the low detection and identification thresholds
for light charged particles (LCPs)
and the very high granularity at forward angles.  
Mass $A$ and charge number $Z$ of detected
reaction products were determined by the energy vs time-of-flight method
(TOF) for LCPs stopped in silicon detectors and by
$\Delta E-E$ ($Z>5$) and
shape identification ($Z \leq 5$) techniques
for charged products stopped in CsI(Tl).
In addition, part of emitted $^8$Be nuclei (two equal-energy $\alpha$'s 
hitting the same crystal) were identified in CsI(Tl)~\cite{Mor10}.
Silicon detectors were calibrated
using proton, carbon and oxygen beams
at various energies ranging from 10 to 100 MeV and energy 
measured with a resolution better than 1\%. 
As $\alpha$-particles of interest lose the major part of their energy in 
CsI(Tl) crystals, a dedicated energy calibration of their fast component 
light was realized using the TOF information and more than 95\% of modules
from 1$^{\circ}$ to 62$^{\circ}$ were calibrated.
The energy resolution for alpha particles varies between 1.0 and 2.5\%
depending on the module.
Further details on $A$ and $Z$ identifications and on the quality of
energy calibrations can be found in
Refs.~\cite{alderighi_nim2002,leneindre_nim2002,adriana_plb}.
In the present work, all reaction products hitting a detection module
were considered as emitted in direction of the geometrical center
of that module
(see~\cite{adriana_plb} for this choice).
When results of simulations were filtered by the multi-detector
replica, the same prescription was applied.
%%%%%%%%%%%%%%%%%%%%% fig.1 %%%%%%%%%%%%%%%%%%%%%%%%%%%%%%% 
\begin{figure}
\begin{center}
\includegraphics[width=0.48\textwidth]{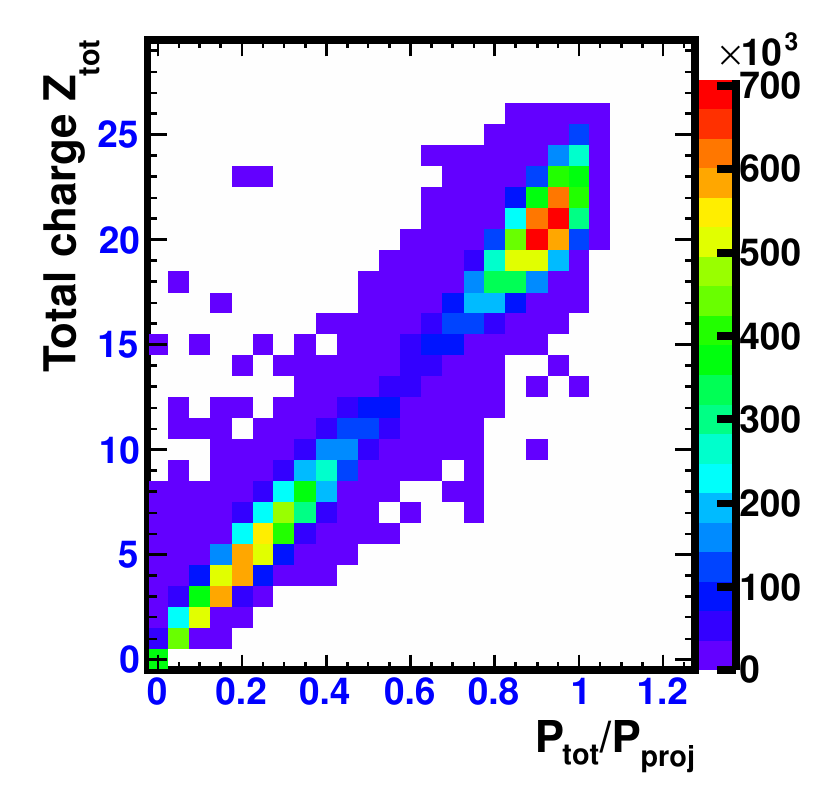}
\includegraphics[width=0.48\textwidth]{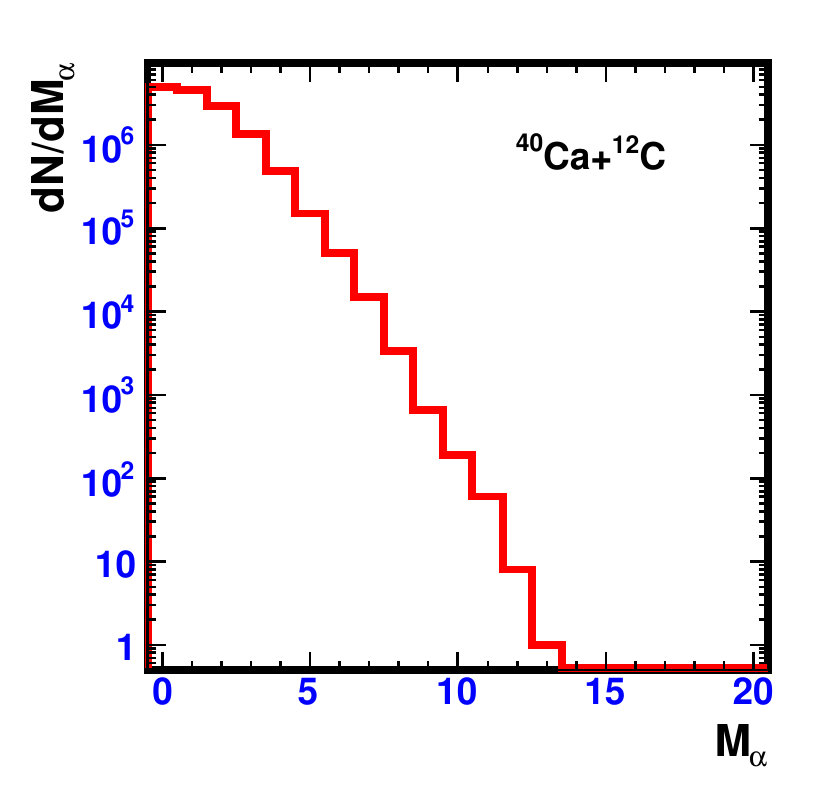}
\end{center}
\caption{(Color online)
Left: contour plot showing event-by-event correlations between
the total detected charge $Z_{tot}$ and the normalized pseudo linear momentum (see text).
Right: distribution of $\alpha$-particle multiplicity, $M_{\alpha}$, for well detected
events ($Z_{tot}\geq$19).}
\label{fig:ztotptot}
\end{figure}
%%%%%%%%%%%%%%%%%%%%%%%%%%%%%%%%%%%%%%%%%%%%%%%%%%%%%%%%%%%%%%%%%%%%%%%%%%%%%%

%%%%%%%%%%%%%%%%%%%selection of events and mechanisms
As a first step in our event selection procedure, we want to exclude
from the data sample poorly-measured events. Without making any
hypothesis about the physics of the studied reaction one can measure
(see Fig. \ref{fig:ztotptot} (left)) the total detected charge $Z_{tot}$
(neutrons are not measured) and the total
pseudo linear momentum normalized to the projectile momentum 
$P_{tot}$/$P_{proj}$ (see Eq.\ref{eq:Zv})
\begin{equation}
P_{tot}=\sum \beta_{par} \gamma Z.
\label{eq:Zv}
\end{equation}
$\beta_{par}$ is the reduced velocity component, with respect to
the beam direction, of the reaction
product of atomic number $Z$ and $\gamma$ is the Lorentz factor.
As the grazing angle of the reaction is 1.11$^{\circ}$, 
to suppress elastic and quasi-elastic reactions, the
first internal ring (1.0$^{\circ}$-1.8$^{\circ}$ polar angle) of CHIMERA
was removed to obtain the data in Fig. \ref{fig:ztotptot} (left).
Well measured events clearly appear in the upper right part of the
figure. In relation with their cross-sections  and with the
geometrical efficiency of CHIMERA, the well detected reaction mechanisms 
correspond to projectile
fragmentation (PF)~\cite{morjean,profrag,LTG} with $Z_{tot}$=19-20
(target fragmentation not
detected) and to incomplete/complete fusion~\cite{sigmafus}
with $Z_{tot}$=21-26. 
At this stage we can have a first indication of the multiplicity
of $\alpha$-particles, $M_{\alpha}$, emitted per event for 
well identified mechanisms ($Z_{tot}\geq$19 - see Fig.
\ref{fig:ztotptot} (right)).
$M_{\alpha}$ extends up to thirteen, which means a
deexcitation of the total system into $\alpha$-particles only. Moreover a
reasonable number of events exhibit $M_{\alpha}$ values up to about 6-7.
%%%%%%%%%%%%%%%%%%%%% fig.2 %%%%%%%%%%%%%%%%%%%%%%%%%%%%%%% 
\begin{figure}
\begin{center}
\includegraphics[width=0.58\textwidth]{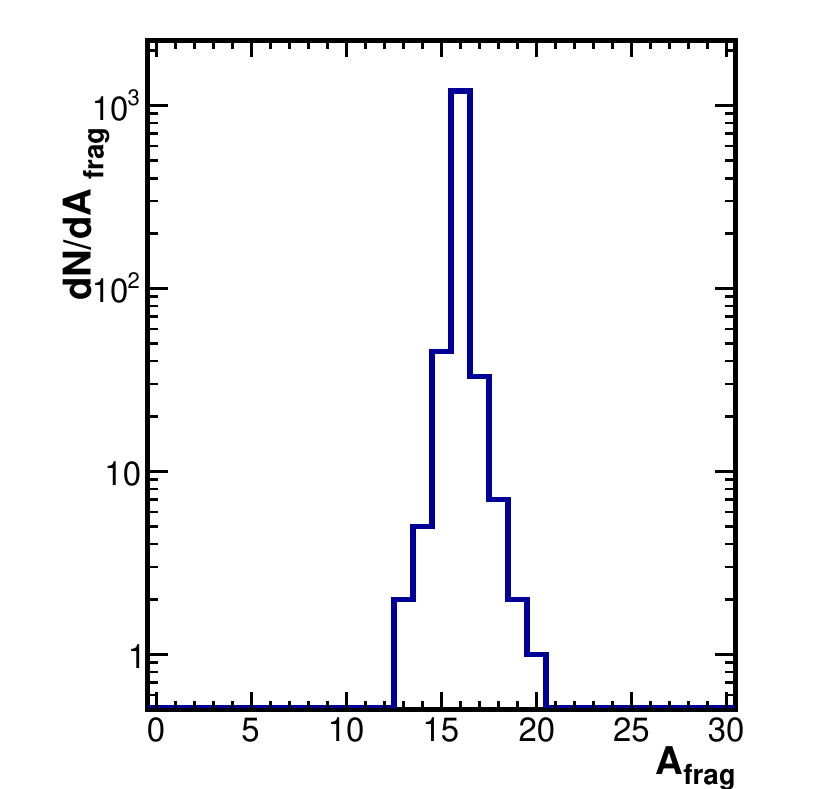}
\end{center}
\caption{(Color online)
Fragment mass number ($A_{frag}$) distribution associated with
selected events ($Z_{tot}$=20) where $Z_{frag}$=8 and $M_{\alpha}$=6.}
\label{fig:distAfrag}
\end{figure}
%%%%%%%%%%%%%%%%%%%%%%%%%%%%%%%%%%%%%%%%%%%%%%%%%%%%%%%%%%%%%%%%%%%%%%%%%%%%%%

The goal is now to tentatively isolate, in events, reaction products emitting
$\alpha$-particles only. Ref.~\cite{morjean} has shown that, at an incident
energy close to ours, $^{20}Ne$ PF is dominated by alpha-conjugate
reaction products. Based on this, and expecting the same for
$^{40}Ca$ PF, we
restrict our selection to completely detected ($Z_{tot}$=20) PF events
composed of one projectile fragment and $\alpha$-particles.
Charge conservation imposes $Z_{frag}$ = 20 - 2$M_{\alpha}$.
Fig. \ref{fig:distAfrag} shows an example of the mass number
distribution, $A_{frag}$, for
$Z_{frag}$=8 associated to $M_{\alpha}$=6. As expected $A_{frag}$=16 largely
dominates; only about 8\% of events correspond to neutron transfers between
projectile and target and lead to an $A_{frag}$ which is different from sixteen.

After this double selection, the question is: from which emission
source are the $\alpha$-particles emitted?
Several possible candidates are present and further
selections must be done before
restricting our study to alpha-sources emitting exclusively the $M_{\alpha}$
observed (called $N\alpha$ sources in what follows).
Possibilities that we shall examine concern: i) pre-equilibrium (PE)
$\alpha$-particle emission,
ii) PF deexcitation through $\alpha$-particle emission proceeding via unbound
states and iii) evaporation from excited Ca projectiles having emitted
$\alpha$-particles only. Concerning deexcitation of PF events via
unbound states, we want, for instance, to exclude from the selection an event composed of
two fragments ($^{24}Mg$ and $^{12}C$*) and one $\alpha$-particle
finally producing one fragment ($^{24}Mg$) and four $\alpha$-particles.

%%%%%%%%%%%%%%%%%%%%% fig.3 %%%%%%%%%%%%%%%%%%%%%%%%%%%%%%% 
\begin{figure}
\begin{center}
\includegraphics[width=0.48\textwidth]{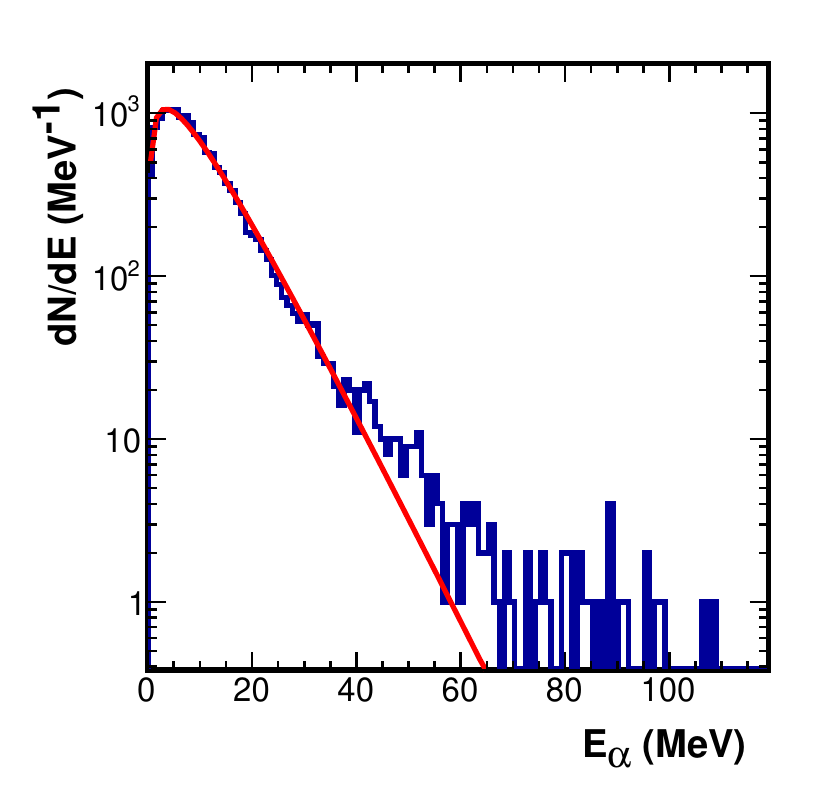}
\end{center}
\caption{(Color online)
$\alpha$-particles emitted by $M_{\alpha}$=5 events: energy spectrum
in the $N_{\alpha}$ reference frame;
curve corresponds to a Maxwellian fit (see text).}
\label{fig:PE}
\end{figure}
%%%%%%%%%%%%%%%%%%%%%%%%%%%%%%%%%%%%%%%%%%%%%%%%%%%%%%%%%%%%%%%%%%%%%%%%%%%%%%

Considering the incident energy of the reaction and the forward
focusing of reaction products, it is important to identify
the possible presence of pre-equilibrium (PE)
$\alpha$-particles~\cite{Lo14,Pa00,Fo14} in
our selected PF events. With the hypothesis that all 
$\alpha$-particles are emitted from their center-of-mass reference frame, 
we can examine the 
corresponding $\alpha$-particle spectra.
Fig. \ref{fig:PE} shows one example of such spectra for $M_{\alpha}$=5.
It exhibits a distribution which resembles a thermal one with the
presence of a high energy tail, which signs PE emission; a similar
spectrum, in the center-of-mass of the reaction, is obtained at
forward angles for the involved collisions.
To prevent errors on alpha emitter properties it is necessary to remove events
in which such PE emission can be present. The curve superimposed on
the spectrum in Fig. \ref{fig:PE}
corresponds to a thermal Maxwellian fit (Coulomb correction, $C_c$, of 0.22 MeV and
temperature, $T$, of 6.5 MeV) with a volume pre-exponential
factor: $dN/dE \propto (E- C_c)^{1/2} exp [-(E-C_c)/T]$~\cite{goldhaber1978}.
It was used to
impose an upper energy limit of 40 MeV, found irrespective of $M_{\alpha}$,
for $\alpha$-particles.
Table \ref{tab:events} displays the percentages of
suppressed events.

%%%%%%%%%%%%%%%%%%%%% fig.4 %%%%%%%%%%%%%%%%%%%%%%%%%%%%%%% 
\begin{figure}
\begin{center}
\includegraphics[width=0.68\textwidth]{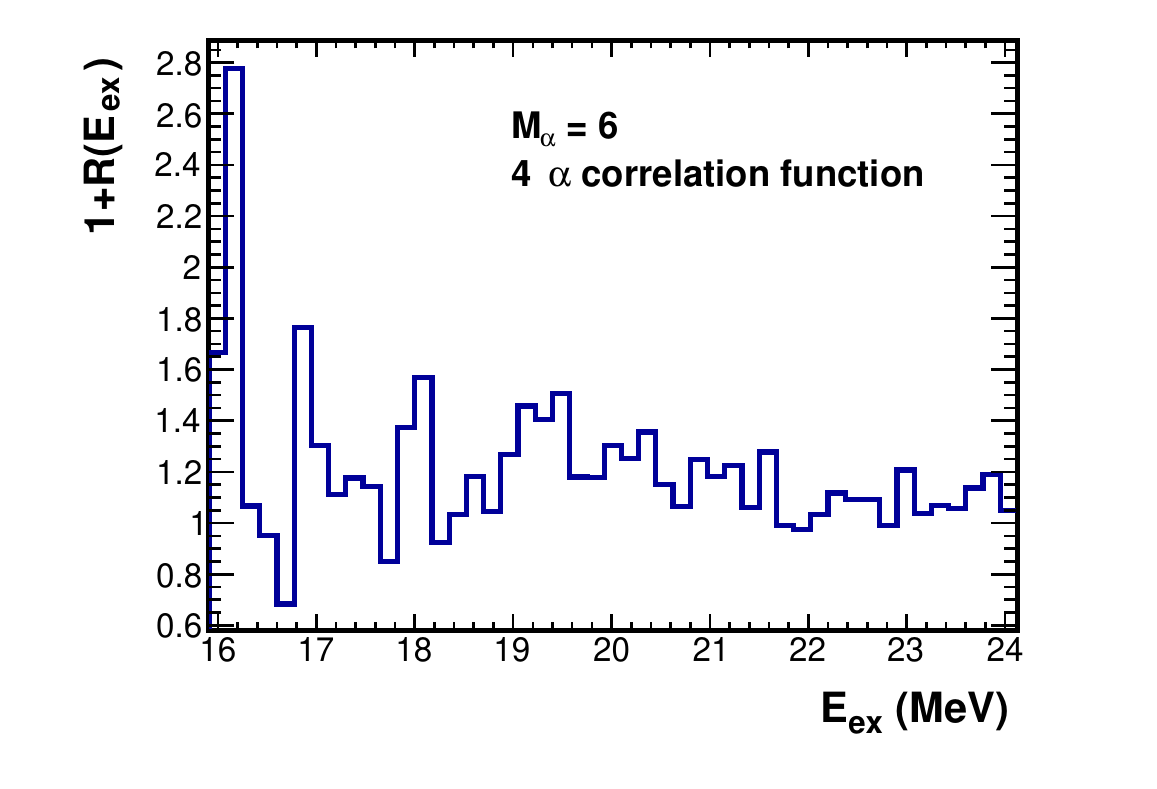}
\end{center}
\caption{(Color online)
$\alpha$-particles emitted by $M_{\alpha}$=6 events:
four $\alpha$-particle correlation function as a function of excitation energy.}
\label{fig:CF}
\end{figure}
%%%%%%%%%%%%%%%%%%%%%%%%%%%%%%%%%%%%%%%%%%%%%%%%%%%%%%%%%%%%%%%%%%%%%%%%%%%%%%

%%%%%%% TABLE 1   percentages removed: preequilibrium and excited% states%%
\begin{table*}[hbt] 
\caption{Information on selected events with different 
$\alpha$ multiplicities, $M_{\alpha}$: percentages of suppressed events
containing pre-equilibrium $\alpha$-particle emission (PE), excited
$\alpha$ sub-systems (see text) and final number of selected events.
Percentages in parentheses correspond to statistical errors.}
\label{tab:events}
\begin{tabular}{p{15mm}p{22mm}|p{22mm}|p{30mm}}
& \multicolumn{1}{c|}{PE $\alpha$-emission} &
\multicolumn{1}{c|}{excited $\alpha$ sub-systems}  \\ \hline
$M_{\alpha}$ & suppressed events (\%) &  suppressed events (\%)
& final number of selected events \\ \hline
 4 & 6.9 (0.2) & 1.6 (0.1) & 12780 \\  
 5 & 7.1 (0.5) & 3.1 (0.3) & 2623 \\
 6 & 9.2 (0.8) & 3.6 (0.5) & 1129 \\  
 7 & 8.7 (1.6) & 3.9 (1.1) & 291 \\  
% 8 & 33.1 (5.2) & 2.5 (1.4) & 78 \\
% 9 & 20.5 (6.8) & 4.6 (2.7) & 33 \\
%10 & 10.3 (5.1) & 7.7 (4.5) & 32 \\
\end{tabular}
\end{table*}
%%%%%%%%%%%%%%%%%%%%%%%%%%%%%%%%%%%%%%%%%%%%%%%%%%%%%%%%%%%%%%%%%%%%%%%%%%%%%%%%%%%%

As far as deexcitation of selected PF events via unbound states is
concerned the use of multi-particle correlation functions (MCFs)~\cite{charity_prc1995}
is required to suppress events.
Correlation function is defined as the ratio between 
the correlated (physical) yield, $Y_{corr}$, 
and the product of single particle yields, generically termed
as uncorrelated spectrum, $Y_{uncorr}$, measured under the same conditions,
\begin{equation}
1+R(X)=\frac{Y_{corr}(X)}{Y_{uncorr}(X)}.
\label{eq:corrf}
\end{equation}
The correlated yield spectrum $Y_{corr}$ is constructed with the
required number of $\alpha$-particles detected in the same event and
we choose to build $Y_{uncorr}$ by mixing particles from different
events~\cite{lisa_prc}. If no correlations are present MCF should
be unity.
The generic variable $X$ is 
the total kinetic energy of the particles of interest in their center-of-mass
frame, $E_{tot}$, or the excitation energy of their emitting source/state, 
$E_{ex}=E_{tot}+Q$; $Q$ is the mass balance. As an example,
for events with $M_{\alpha}$=6 we build
three-, four- and five-particle MCFs considering all possible
combinations of $\alpha$-particles. Fig. \ref{fig:CF} displays the
four-particle MCF. Peaks, statistically significant, 
are located in the excitation energy range 16.7 - 22.0 MeV.
Considering the experimental resolution measured,
around 350 keV~\cite{adriana_plb}, the following peaks or sum of two
peaks which correspond
to unbound states of $^{16}O$ are observed:
16.84 + 17.20, 17.72 + 18.09, 19.26 + 19.54, 20.05 + 20.41, 21.05 and
21.65 MeV;
%Several peaks which
%correspond to unbound states of $^{16}O$
%(15.83, 16.27, 16.84, 17.72, 18.09, 18.78, 19.26 MeV {\ldots}) are observed;
some of them 16.84 and 21.05 being known as 100\% $\alpha$-particle
emitters. This is a clear indication that
some deexcitation via those states occurs.
From those MCF studies it appears that only two  nuclei, $^{12}C$ (Hoyle state and
broad peak centered at 9.64 MeV excitation energy) and $^{16}O$
(see Fig. \ref{fig:CF})
significantly contribute to deexcitation via their unbound states.
For each $M_{\alpha}$ value, average MCF values corresponding to each state have been
calculated and only a percentage of events (from 1 to 95 \%) with
$\alpha$-particles populating those states were
kept. Percentages kept correspond to the weights (in percent) of background
levels under the peaks: 1 - [(MCF-1)/MCF] = 1/MCF.
Table~\ref{tab:events} shows the small percentages of
suppressed events (from 1.6 to 3.9 \%), which obviously concern events
with low energy $\alpha$-particles in their $\alpha$ reference frame.
Final numbers of selected events for each of the
$M_{\alpha}$ values are also indicated, which show sufficient
statistics for comparisons with simulations for $M_{\alpha}$=4-6.

\begin{figure}
\begin{center}
\includegraphics[width=0.48\textwidth]{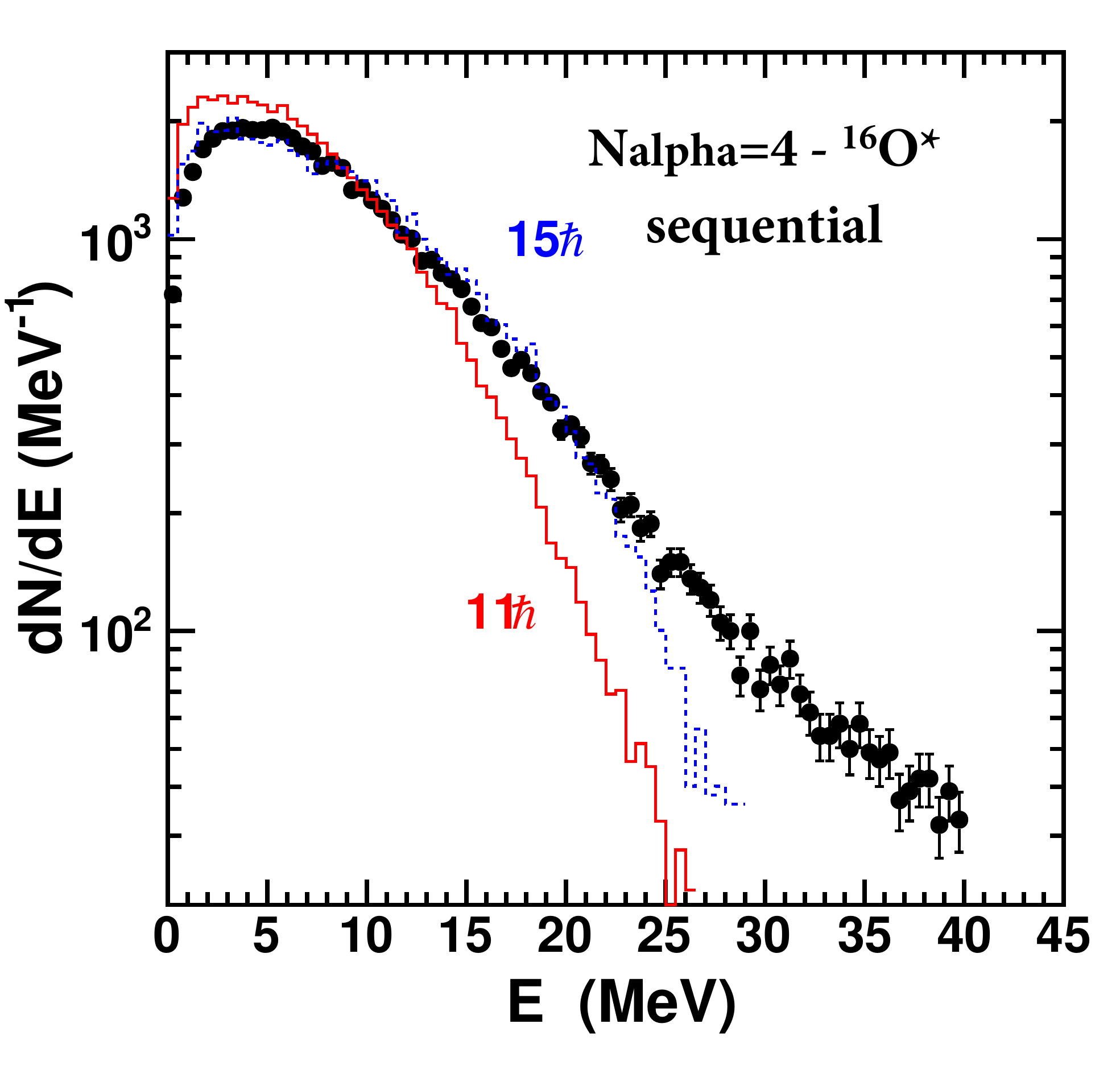}
\includegraphics[width=0.48\textwidth]{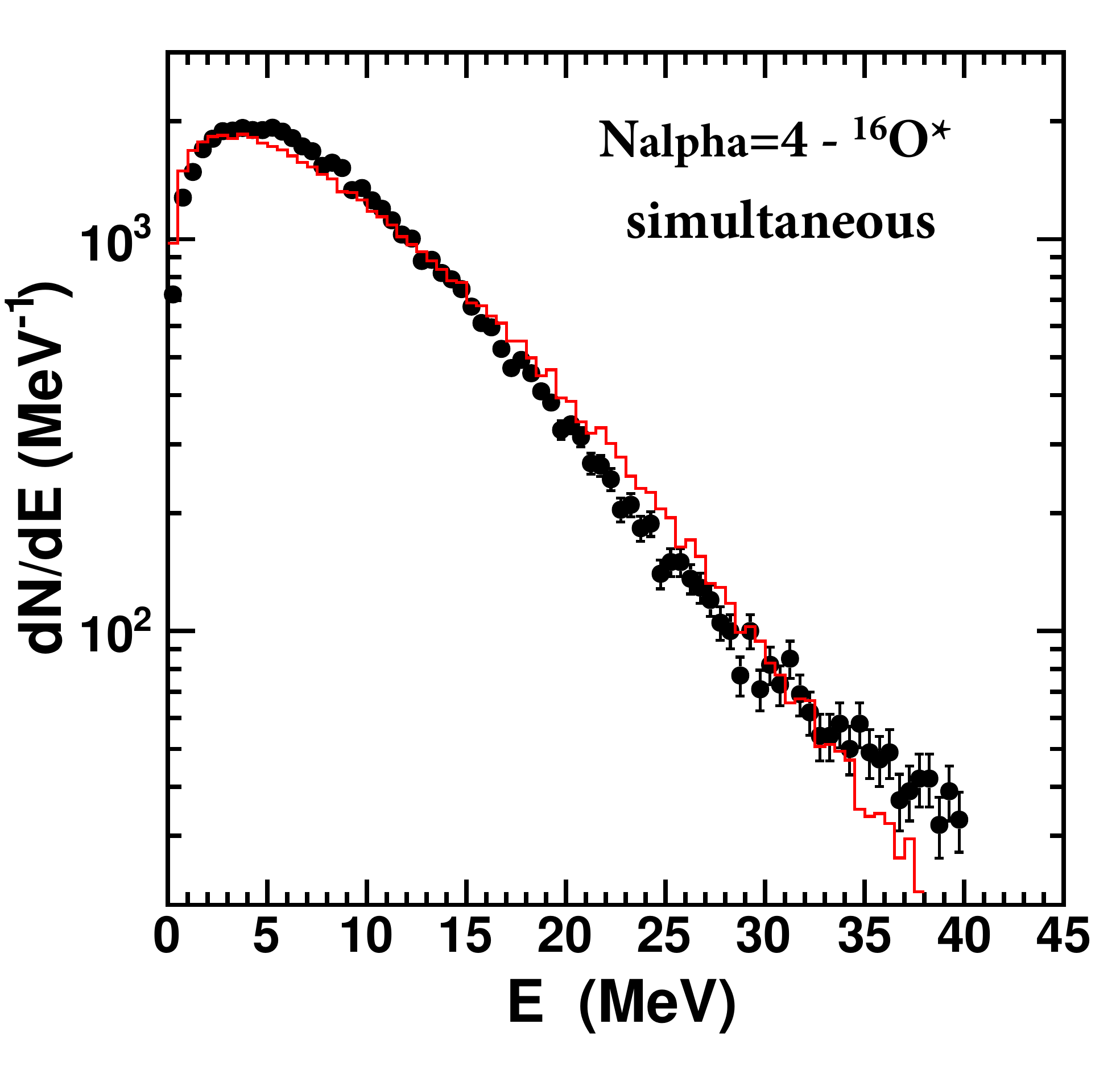}

\includegraphics[width=0.48\textwidth]{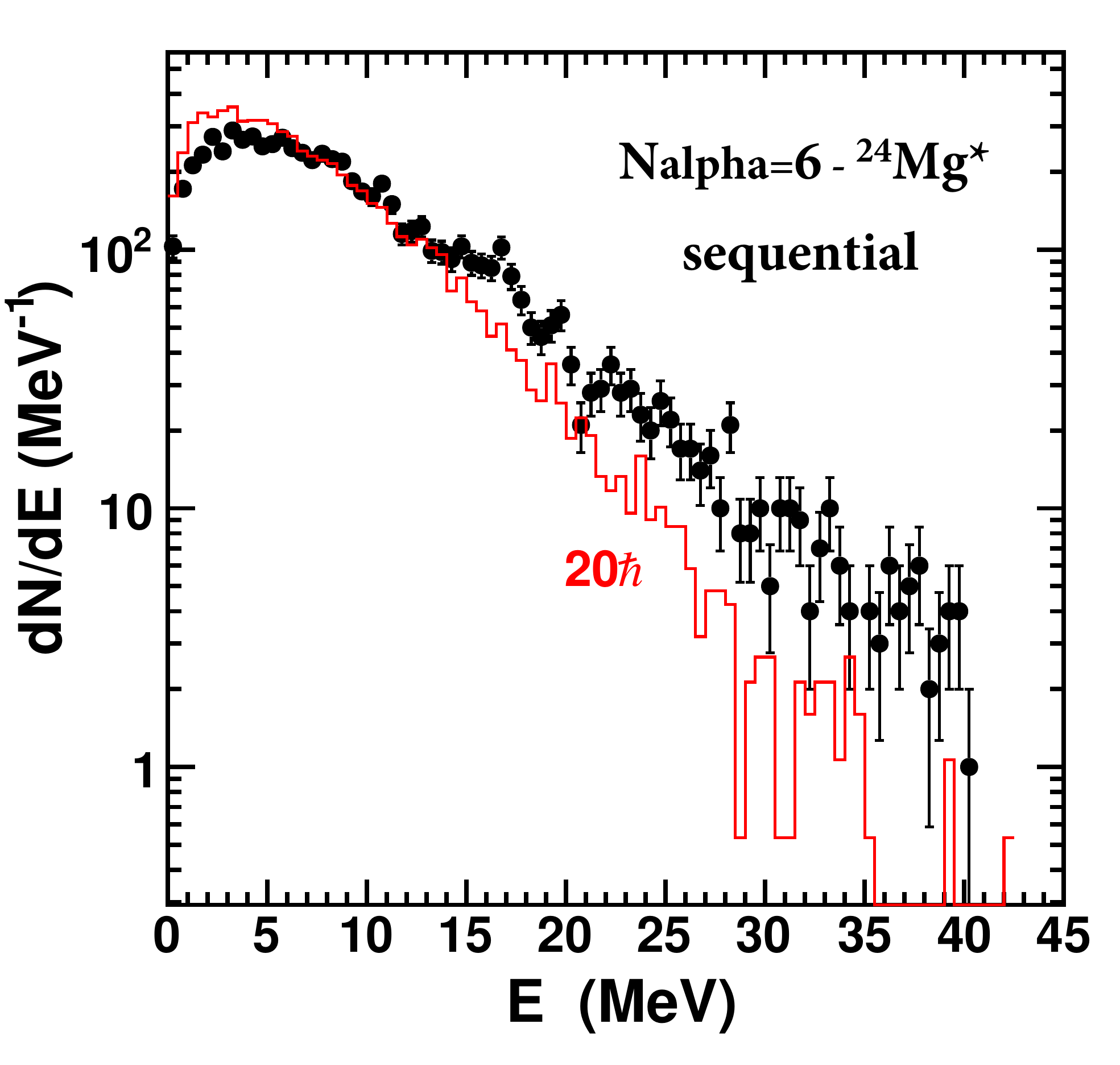}
\includegraphics[width=0.48\textwidth]{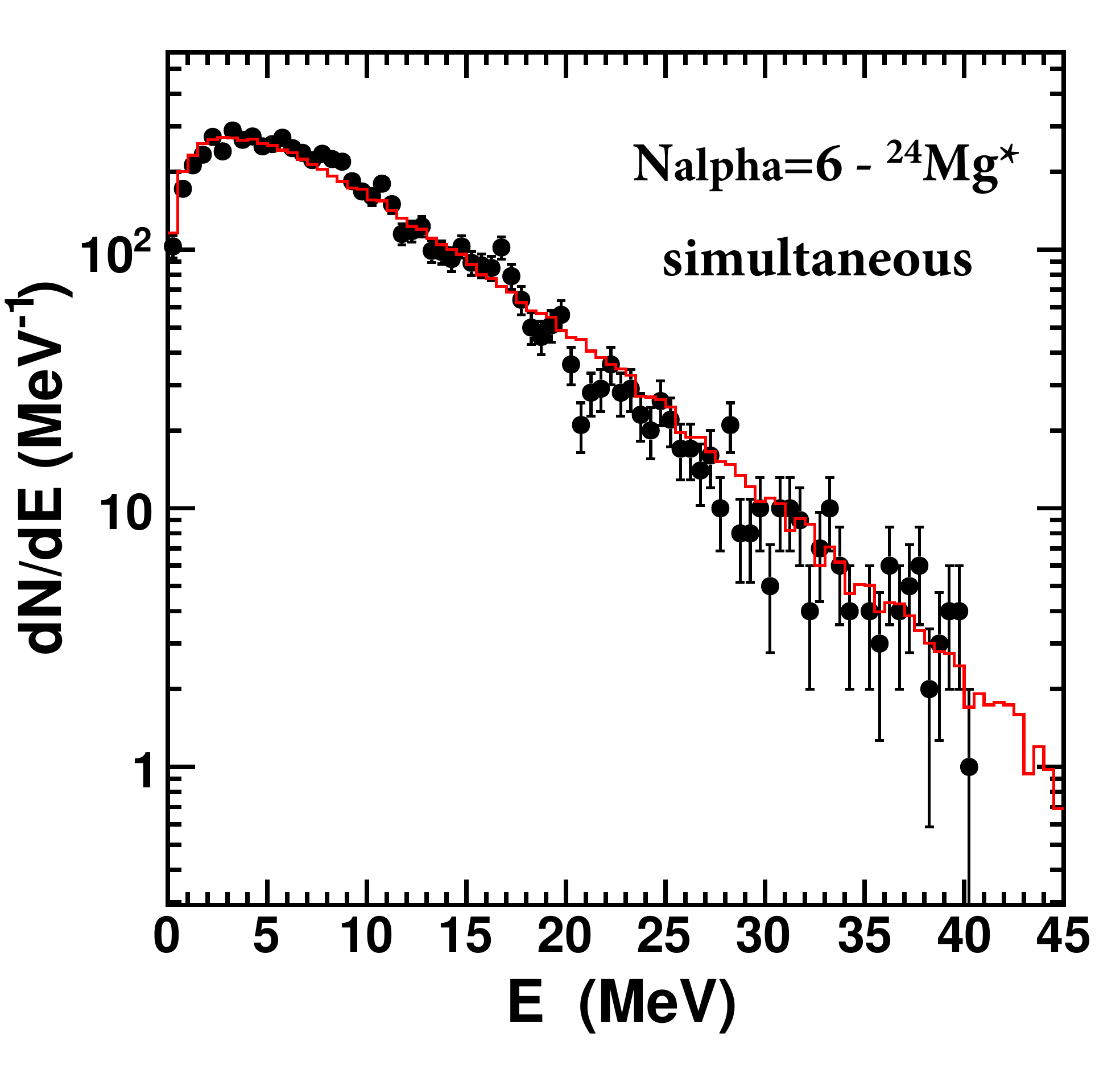}
\end{center}
\caption{(Color online)
Particle spectra from $N\alpha$ sources: $^{16}O$*, top and 
$^{24}Mg$*, bottom; black dots with
statistical error bars correspond to experimental data. Histograms
superimposed on data correspond to filtered simulations: left - for
sequential decay and different spin distributions with
GEMINI++ and right - with a dedicated simulation of a simultaneous decay
process (see text).}
\label{fig:EXEA4}
\end{figure}

As far as $N\alpha$ sources are concerned,
the effect of suppressing events is to 
reduce both the mean values and the widths of their excitation energy distributions.
As an example, for $N\alpha$=4, mean excitation decreases from 56.2 to
52.4 MeV and the root mean square (RMS) of the distribution
from 22.7 to 15.7 MeV. The corresponding
$\alpha$-energy spectrum
(top, black dots) is displayed
in Fig.~\ref{fig:EXEA4}; for comparison same information is also
reported for $N\alpha$=6 (bottom, black dots). For deduced excitation
energies we did not consider
deexcitations via $^8Be$ because they do not significantly modify the
conclusions. However, for completeness, Table~\ref{tab:8Be} 
shows the percentages of one and two $^8Be$ emissions measured
from relative energy spectra of two $\alpha$-particles and
associated correlation functions for different $M_{\alpha}$ values; percentages with three
$^8Be$ are found negligible ($<1$\%).
Information relative to $^8Be$ emission  will be important to discuss sequential versus
simultaneous $\alpha$-emission from $N\alpha$ sources.
%We recall that events containing $^8Be$ nuclei 
% for which the two $\alpha$-particles  
%hit the same detection module were identified in CsI(Tl) crystals and not selected.
%%%%%%%%%%%%%%%%%%%%% fig.5 %%%%%%%%%%%%%%%%%%%%%%%%%%%%%%% 
%%%%%%%%%%%%%%%%%%%%%%%%%%%%%%%%%%%%%%%%%%%%%%%%%%%%%%%%%%%%%%%%%%%%%%%%%%%%%%

To conclude on this part, one can indicate that if excited $N\alpha$
sources have been formed their excitation energy
thresholds for total deexcitation into $\alpha$-particles vary from 20
to 50 MeV as $N\alpha$ varies from 4 to 6. Their mean excitation
energy per nucleon is rather constant around 3.3-3.5 MeV. 
One can also deduce a crude estimation of
the average lowest density they may have reached due to thermal pressure
before decaying into $\alpha$-particles.
To do that, starting from a phenomenological quadratic equation of
state (EoS)~\cite{stocker86},
an EoS for subnormal density of finite nuclear systems 
%requiring a binding energy of 8 MeV at normal density and going to zero when the
%density vanishes
was proposed in~\cite{friedman_prc}:
\begin{equation}
(E/A)_{T=0}=8 [(1-\rho / \rho_0)^2 -1] MeV.
\label{eq:density}
\end{equation}
From Eq. \ref{eq:density} introducing an initial excitation energy
at normal density and ignoring any dissipative processes during the
expansion stage
(excitation energy = expansion energy) an estimation of the minimum density reached
can be calculated~\cite{borderie_brolo}.
The minimal average density $\rho$ which was derived is found around
0.7$\rho_0$ where $\rho_0$ is the normal density.

%%%%%%% TABLE 2   deexcitation via 8Be avec GEMINI %%%%%%%%%%%%%%%%%%%%%%%%%%%%%%%%%%%%
\begin{table*}[hbt] 
\caption{Percentages of selected events which deexcite via one or two
$^8Be$ as a function of $\alpha$ multiplicity $M_{\alpha}$;
percentages in parentheses correspond to statistical errors. Results
from GEMINI++ simulations for $N\alpha$ sources are also reported.}   
\label{tab:8Be}
\begin{tabular}{p{10mm}p{22mm}p{22mm}|p{22mm}p{30mm}}
& \multicolumn{1}{c}{selected events} &
\multicolumn{3}{c}{GEMINI++ results}  \\ \hline
$M_{\alpha}$ & one $^8Be$  &  two $^8Be$  & one $^8Be$ &  two $^8Be$ \\ \hline
 4 & 7.7 (0.3)  &   0.2      & 100.0  &  0.0\\  
 5 & 12.0 (0.7) &   0.2      & 59.4  &  40.6\\
 6 & 19.3 (1.3)&   0.4        & 13.5  &  86.5\\
% 7 & 24.0 (2.7)&   2.7       &  7.8  &  87.7 \\  
% 8 & 24.6 (4.5)&   2.5 \\
% 9 & 38.8 (9.4)&   9.1 \\
%10 & 53.1 (11.7)&  15.4 \\
\end{tabular}
\end{table*}
%%%%%%%%%%%%%%%%%%%%%%%%%%%%%%%%%%%%%%%%%%%%%%%%%%%%%%%%%%%%%%%%%%%%%%%%%%%%%%%%%%%%

Before discussing different possible deexcitations involved for
final retained events,
information on projectile fragmentation mechanism is needed. 
%For the considered reaction a grazing angular momentum
%$l_{max}$=90$\hbar$ is calculated~\cite{sigmar}. Assuming a sharp triangular angular
%momentum distribution, $l_{fus}$=35$\hbar$ (total fusion) and
%$l_{CF}$=24$\hbar$ (complete fusion)
%can be estimated~\cite{sigmafus}.
Global features of PF events 
are reproduced by a model of stochastic transfers~\cite{LTG}. Main
characteristics for
primary events with $Z_{tot}$=20 are the following: i) excitation energy
 extends up to about 200 MeV, which 
allows the large excitation energy domain
(20-150 MeV) measured for $N\alpha$ sources when associated to a single
fragment, 
and ii) angular momenta extend up to 24$\hbar$, which gives
an upper spin limit for excited Ca projectiles and consequently
for $N\alpha$ sources.

Are $\alpha$-particles emitted sequentially or simultaneously?
To answer the question $\alpha$-energy spectra can be compared to
simulations. For excited Ca projectiles and $N\alpha$ sources,
experimental velocity and excitation
energy distributions as well as distributions for spins are used as
inputs. Results of simulations are then
filtered by the multi-detector replica including all detection and
identification details. 
%$N\alpha$ sources are finally recontructed
%and related energy spectra compared to data;
Simulated spectra are
normalized to the area of experimental spectra. 

For sequential emission the GEMINI++ code~\cite{gemini,TST} was used.
It combines the Hauser-Feshbach formalism for evaporation of
particles ($Z<$ 5)
and the transition-state formalism for intermediate-mass fragment
emission ($Z\geqslant$ 5).
Evaporation includes $n, p, d, t,$ $^3He$, $\alpha$-particles,
$^6He$, $^{6-8}Li$ and $^{7-10}Be$ channels.
For fragment emission, the saddle conditional energy for
different mass (or charge) asymmetries is deduced from the finite
range rotating liquid-drop model~\cite{sierk}. Note that emitted
$^8Be$ are directly transformed into two $\alpha$-particles.

\begin{figure}
\begin{center}
\includegraphics[width=0.48\textwidth]{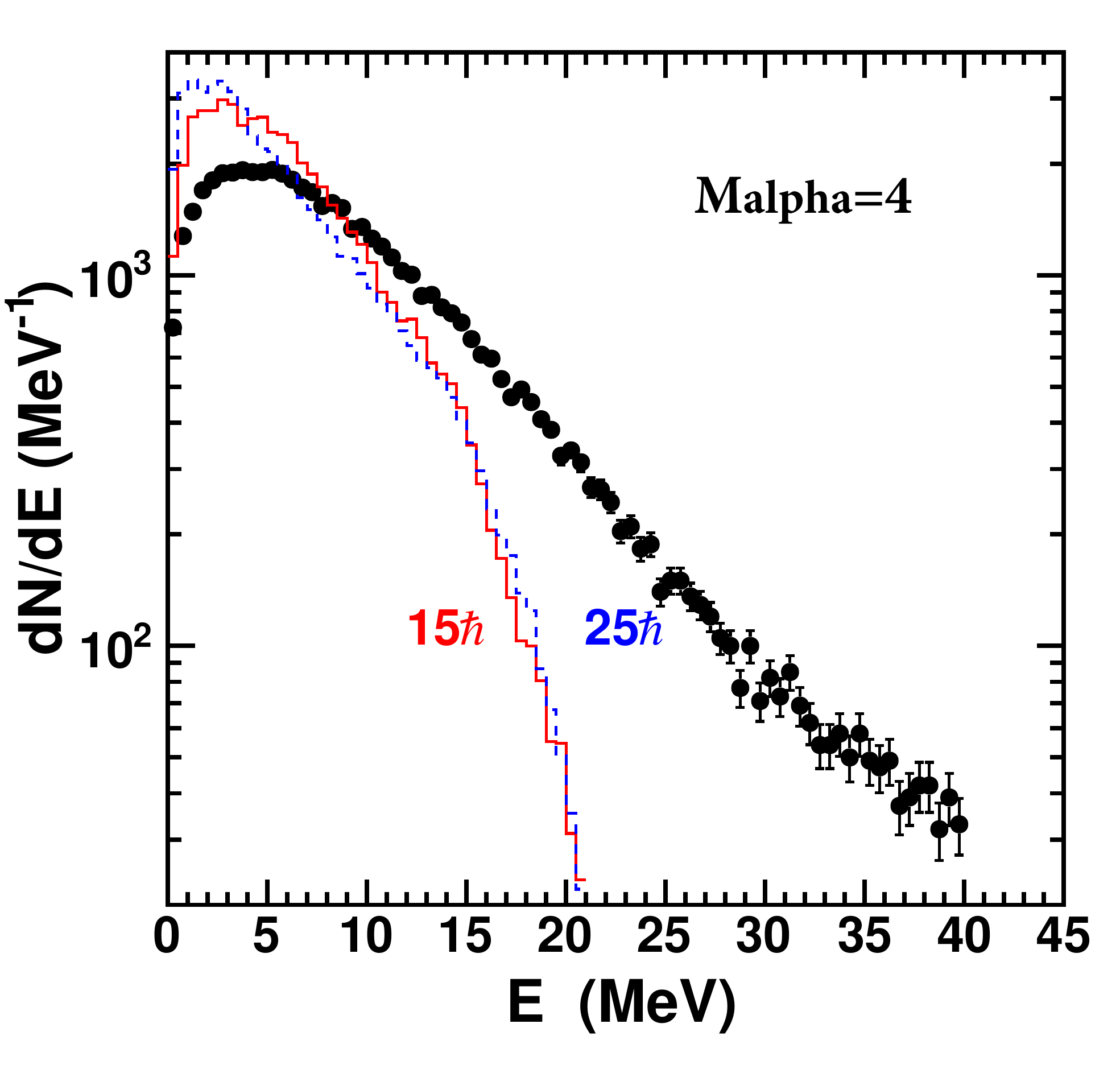}
\includegraphics[width=0.48\textwidth]{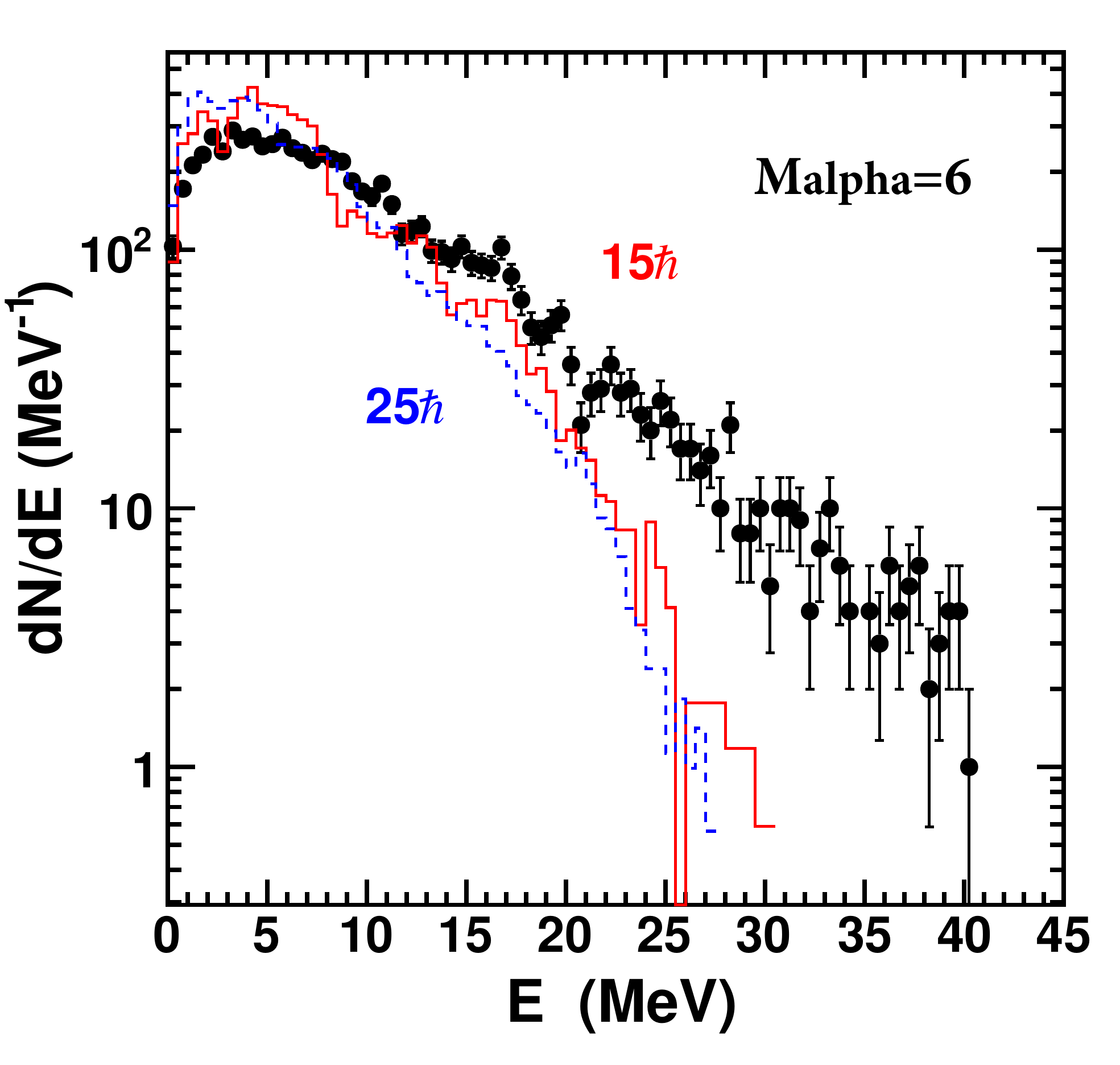}
\end{center}
\caption{(Color online)
Sequential decay of excited Ca projectiles: 
	energy spectra (in the $M_{\alpha}$=4 (left) or $M_{\alpha}$=6 (right) system
	reference frame) of
	evaporated $\alpha$-particles associated to a $^{24}Mg$ (left)
	or to an $^{16}O$ (right)
	evaporation residue. Black dots are experimental data and
	histograms are results of GEMINI++ simulations (see text).
}
\label{fig:SEQPF}
\end{figure}

Before discussing decays of $N\alpha$ sources, we must consider 
the possible evaporation from Ca projectiles as stated previously.
Excitation energy for projectiles is deduced from 
%Eq.\ref{eq:exQP}
$E^*$=$E^{*}(N\alpha)$+$E_{rel}$+$Q$.
%\begin{equation}
%E^*=E^{*}(N\alpha) + E_{rel} + Q.
%\label{eq:exQP}
%\end{equation}
$E_{rel}$ is the relative energy between the $N\alpha$ source and the
associated fragment (evaporation residue). Fig. \ref{fig:SEQPF} (left)
displays, for $M_{\alpha}$=4, results of simulations (histograms)
with reconstructed excitation energy distribution for
$^{40}Ca$ ($<E^*>$=68.8 MeV) and gaussian distributions centered at 15
and 25$\hbar$ for spins as inputs.
$M_{\alpha}$=4 and $^{24}Mg$ residues are the decay products after filtering.
A similar comparison is displayed in Fig. \ref{fig:SEQPF} (right) for  
$M_{\alpha}$=6; excitation energy distribution for
$^{40}Ca$ is centered at $<E^*>$=99.7 MeV.
RMSs used for spin distributions, around 2.0-2.5$\hbar$, were deduced from the correlation
``excitation energy-spin'' obtained with the model of stochastic transfers~\cite{LTG}.
Note that no more $^{24}Mg$, $^{20}Ne$ or $^{16}O$  residues are
produced in simulations for $^{40}Ca$ spin distributions
centered at values larger than 25$\hbar$.
Comparisons with experimental data (full points) show a poor agreement indicating that
such an hypothesis of sequential decay from excited projectiles
is not correct. The same conclusion is derived for
$M_{\alpha}$ equal 5~\cite{bb_nn15}. 
%%%%%%%%%%%%%%%%%%%%% fig.6 %%%%%%%%%%%%%%%%%%%%%%%%%%%%%%% 
%%%%%%%%%%%%%%%%%%%%%%%%%%%%%%%%%%%%%%%%%%%%%%%%%%%%%%%%%%%%%%%%%%%%%%%%%%%%%%

Considering sequential deexcitation of $N\alpha$ sources,
histograms in Fig.~\ref{fig:EXEA4} (left) are examples of GEMINI++ simulation
results for $N\alpha$=4 ($^{16}O$*) and for $N\alpha$=6 ($^{24}Mg$*).
Gaussian distributions for spins are used as
inputs and the best agreement with data is obtained with RMS=1.5$\hbar$
for spin distributions.
The agreement between 
data and simulations becomes progressively poorer as the $N\alpha$ value decreases.
%For $N\alpha$=7 a rather good agreement between 
%data and simulations is obtained whereas it 
%appears poorer and poorer when $N\alpha$ value decreases.
%Note that for $N\alpha$ larger than seven
%it was not possible (with 200000 simulated GEMINI deexcitations)
%to obtain sequential decays with exclusively $\alpha$-particles.
However, the most important disagreement
between data and simulations
concerns the percentages of $N\alpha$ sources which
deexcite via $^8Be$ (see Table~\ref{tab:8Be}).
%A sequential $\alpha$-emission from $N\alpha$ systems 
%ends with the last evaporation step leaving an unstable
%$^8Be$ residue, which means one $^8Be$ emission per event. 
With the excitation energy distributions experimentally deduced the
GEMINI++ code evaporates an important percentage of
$^8Be$ along the deexcitation chain and at
the last evaporation step of the chain leaving an unstable
$^8Be$ residue.
This strong constraint is not experimentally observed.

For simultaneous emission from $N\alpha$ sources, a dedicated simulation was done which
mimics a situation in which $\alpha$ clusters are early formed when the
$N\alpha$ source is expanding~\cite{girod_prl2013,ebran_2014} due to thermal
pressure.
By respecting the experimental excitation energy distributions of 
$N\alpha$ sources, a distribution of $N\alpha$ events is generated as
starting point of the simulation.
Event by event, the $N\alpha$ source is first split into $\alpha$'s.
Then the remaining available energy ($E^* + Q$) is directly randomly
shared among the $\alpha$-particles such as to conserve energy and linear
momentum~\cite{lopez}.
Histograms in Fig. \ref{fig:EXEA4} (right) are the results of such a simulation,
which show a good agreement with data. 
Similar calculated energy spectra 
were also obtained with simulations containing an intermediate freeze-out
volume stage where $\alpha$-particles are formed and then propagation of particles
in their mutual Coulomb field.
In this case angular momentum distributions of $N\alpha$ sources at
freeze-out
can also be deduced: they exhibit a Maxwell-like shape extending up
to 25$\hbar$ for $N\alpha$=6 while mean values vary from 6.7 to 9.5$\hbar$
when $N\alpha$ moves from 4 to 6.
Note that $^8Be$ emission is out of the scope of the present
simulation. 

From these comparisons with both sequential and
simultaneous decay simulations it clearly appears that sequential
emission is not able to reproduce experimental data whereas a remarkable
agreement is obtained when an
$\alpha$-clustering scenario is assumed. Same conclusion is derived
for $N\alpha$ equal 5~\cite{bb_nn15}.
However one cannot exclude that a small
percentage of $N\alpha$ sources, those produced with lower excitation
energies, sequentially deexcite.

In conclusion, the reaction $^{40}$Ca+$^{12}$C at 
25 MeV per nucleon bombarding energy was used to produce and
carefully select specific classes of projectile fragmentation events from which excited
$N\alpha$ sources can be unambiguously identified.
Their excitation energy distributions are derived with mean values
around 3.4 MeV per nucleon and a crude estimation of their mean minimal
densities, around 0.7 the normal density, can be deduced.

%, which indicates that mean densities smaller than normal density
%have been reached.
Their energetic emission properties were compared with two simulations,
one involving sequential decays and a second for simultaneous decays.
For excited expanding $N\alpha$ sources composed of 4, 5 and 6 $\alpha$-particles,
for which statistics is good enough for conclusive comparisons with
simulations,
evidence in favor of simultaneous emission
($\alpha$-particle clustering) is reported. Those results support mean
field calculations of Refs.~\cite{girod_prl2013,ebran_2014}.

%%%%%%%%%%%%%%%%%% Biblio %%%%%%%%%%%%%%%%%%%%%%%%%%%%%%%%%%%%%%%%%%%%%%%

\end{document}